\documentclass{article}

\usepackage{arxiv}

\usepackage[utf8]{inputenc} % allow utf-8 input
\usepackage[T1]{fontenc}    % use 8-bit T1 fonts
\usepackage[hidelinks]{hyperref}       % hyperlinks
\usepackage{url}            % simple URL typesetting
\usepackage{booktabs}       % professional-quality tables
\usepackage{amsfonts}       % blackboard math symbols
\usepackage{amsmath}
\usepackage{nicefrac}       % compact symbols for 1/2, etc.
\usepackage{microtype}      % microtypography
\usepackage{lipsum}
\usepackage{graphicx}
\usepackage{multirow}
\graphicspath{ {./images/} }

\title{Solar Active Region Magnetogram Image Dataset for Studies of Space Weather
\thanks{\textit{\underline{Citation}}: 
\textbf{Boucheron, L.E., Vincent, T., Grajeda, J.A., and Wuest, E. Solar active region magnetogram image dataset for studies of space weather. Sci Data 10, 825 (2023). https://doi.org/10.1038/s41597-023-02628-8}}}

\author{
 Laura E.~Boucheron \\
  Klipsch School of Electrical and Computer Engineering\\
  New Mexico State University\\
  Las Cruces, NM 88001, USA \\
  \texttt{lboucher@nmsu.edu} \\
   \And
 Ty Vincent \\
  Klipsch School of Electrical and Computer Engineering\\
  New Mexico State University\\
  Las Cruces, NM 88001, USA \\
  \And
 Jeremy A.~Grajeda \\
  Klipsch School of Electrical and Computer Engineering\\
  New Mexico State University\\
  Las Cruces, NM 88001, USA \\
  \texttt{jgra@nmsu.edu} \\
  \And
  Ellery Wuest \\
  Klipsch School of Electrical and Computer Engineering\\
  New Mexico State University\\
  Las Cruces, NM 88001, USA \\
  \texttt{ellerywu@nmsu.edu} \\
}

\begin{document}
\maketitle
\begin{abstract}
In this dataset we provide a comprehensive collection of magnetograms (images quantifying the strength of the magnetic field) from the National Aeronautics and Space Administration's (NASA's) Solar Dynamics Observatory (SDO).  The dataset incorporates data from three sources and provides SDO Helioseismic and Magnetic Imager (HMI) magnetograms of solar active regions (regions of large magnetic flux, generally the source of eruptive events) as well as labels of corresponding flaring activity.  This dataset will be useful for image analysis or solar physics research related to magnetic structure, its evolution over time, and its relation to solar flares.  The dataset will be of interest to those researchers investigating automated solar flare prediction methods, including supervised and unsupervised machine learning (classical and deep), binary and multi-class classification, and regression.  This dataset is a minimally processed, user configurable dataset of consistently sized images of solar active regions that can serve as a benchmark dataset for solar flare prediction research. 
\end{abstract}

% keywords can be removed

\section{Background \& Summary}
In this dataset, we provide a comprehensive collection of magnetograms (images quantifying the strength of the magnetic field) from the National Aeronautics Space Administration's (NASA's) Solar Dynamics Observatory (SDO).  SDO was launched on 11 February 2010 as the first mission in support of the Living With a Star (LWS) program which seeks to understand solar variability and the effects of space weather at Earth and throughout the Solar System~\cite{pesnell2012}.  Specific goals of SDO in line with this dataset are to better understand the magnetic structure of the Sun and understand and predict how that magnetic structure initiates space weather events such as flares~\cite{pesnell2012}. Three experiments are included on SDO: the Atmospheric Imaging Assembly (AIA)~\cite{lemen2012}, the EUV Variability Experiment (EVE)~\cite{woods2012}, and the Helioseismic and Magnetic Imager (HMI)~\cite{scherrer2012}.  In this paper, we focus on line-of-sight magnetogram images from HMI.  

The dataset presented in this paper provides a comprehensive set of HMI magnetograms of solar active regions (regions of large magnetic flux, generally the source of eruptive events) as well as labels of corresponding flaring activity.  This dataset will be useful for image analysis or solar physics research related to magnetic structure, its evolution over time, and its relation to solar flares (a sudden and large emission of radiation).  It is expected that the main audience for this dataset are those researchers investigating automated solar flare prediction methods, including supervised and unsupervised machine learning (classical and deep), binary and multi-class classification, and regression.  While SDO provides an incredibly rich dataset that can be an excellent source for image processing and machine learning researchers, there are several characteristics of the data that motivated our creation of this specific dataset.  First, and overarching, was the desire to provide a minimally processed, user configurable dataset that can serve as a benchmark dataset for solar flare prediction research.  Second was the desire to focus analysis on solely active regions and to reduce the amount of time needed to interact with existing interfaces to download such data.  Third was the desire that images of those active regions be consistently sized images rather than varying across active regions and/or across time.  Fourth was the necessity of integrating a separate dataset in order to develop labels related to flare activity.  

In this dataset, we address the aforementioned characteristics as follows.  First, we provide a comprehensive set of magnetogram images from all National Oceanic and Atmospheric Administration (NOAA) active regions (ARs) from May 2010 through December 2018.  Along with this set of images, we provide a means to configure basic parameters of the dataset, including the size of flares to consider, the time window over which to consider flare prediction, the latitudes and longitudes of active regions to include, and whether to include images with Not-a-Number (NaN) pixel values.  Second, we integrate two sources of data in order to retrieve data only associated with ARs and provide a means to automate the download of those AR magnetogram images.  Third, we provide consistently sized ($600\times600$ pixel) images, which can be an important assumption in batch processing of images, particularly for some common deep learning methods, e.g., convolutional neural networks.  Fourth, we integrate a third source of data in order to provide labels related to flaring activity.  

\section{Methods}
\subsection{Dataset Overview}
\begin{figure}[t]
    \centering
    \includegraphics[width=\textwidth]{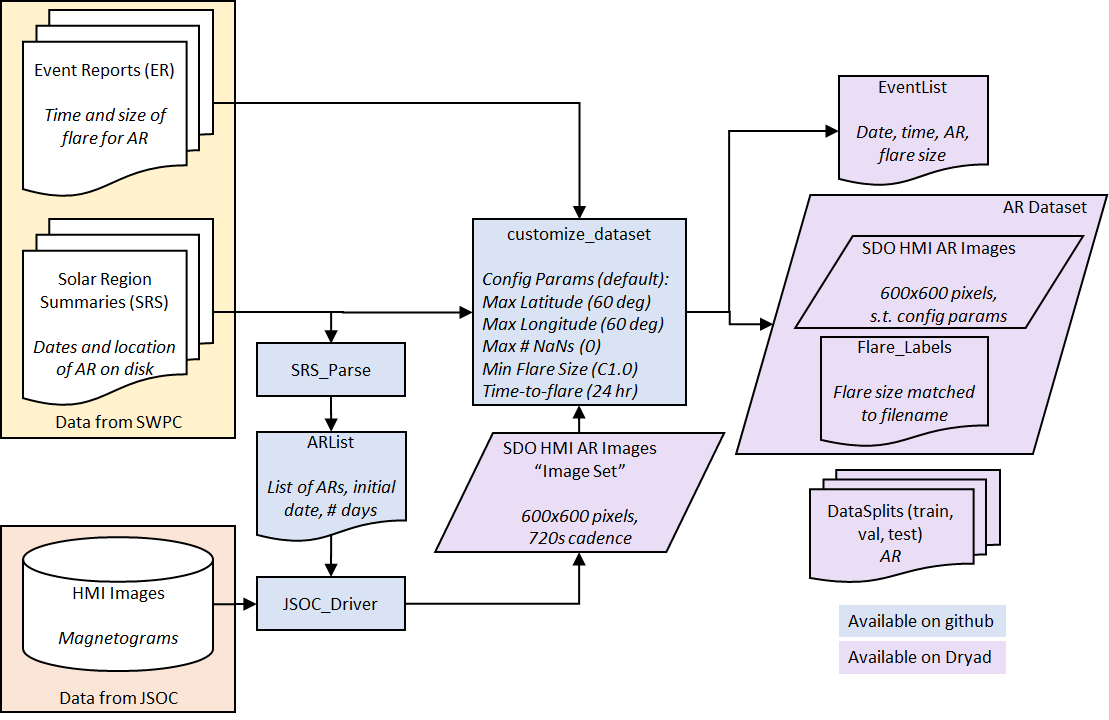}
    \caption{Flowchart of dataset creation.  Space Weather Prediction Center (SWPC) Solar Region Summaries (SRS) are used to determine the dates for which a National Oceanic and Atmospheric Administration (NOAA) Active Region (AR) is visible on disk.  Solar Dynamics Observatory (SDO) Helioseismic and Magnetic Imager (HMI) magnetogram images of ARs are downloaded via the Joint Science Operations Center (JSOC) web interface.  SWPC Event Reports (ER) are used to specify the time and size of solar flares associated with a given NOAA AR.}
    \label{fig:data_flowchart}
\end{figure}

This dataset incorporates data from three main sources. First, in order to focus the image collection on ARs, we used the NOAA Space Weather Prediction Center (SWPC) Solar Region Summaries (SRS) (\url{ftp://ftp.swpc.noaa.gov/pub/warehouse/}) and parsed those text data to extract the date an AR appeared on disk and the number of days it was visible on disk.  Additionally, the SRS provide latitude and longitude of ARs which we use to postprocess the dataset.  Second, we download magnetogram images from SDO/HMI using the Joint Science Operations Center (JSOC) interface (\url{http://jsoc.stanford.edu/ajax/lookdata.html}) at a cadence of 720 seconds, centered at the NOAA AR centroid (tracked according to the Carrington rate), and with a spatial extent of $600\times600$ pixels.  This image size was chosen to correspond to approximately 300 arcseconds $\times$ 300 arcseconds ($300''\times300''$) commensurate with previous work on solar flare prediction, e.g.,~\cite{algraibah2015,boucheron2015}, and to be large enough to encompass the typical range of AR sizes~\cite{canfield2000}.  Third, we used the SWPC Event Reports (ER) (\url{ftp://ftp.swpc.noaa.gov/pub/warehouse/}) to extract the AR number, peak flare time, and flare size in order to provide labels for those researchers investigating a supervised classification or regression problem. Figure~\ref{fig:data_flowchart} summarizes the data flow used to create this dataset.

In total, we downloaded images corresponding to 1,655 NOAA ARs which appeared with sunspot structure on the Sun from 01 May 2010 through 31 December 2018, a total of 1,372,004 HMI images from NOAA ARs 11064 through 12731. We only include those ARs which appeared for the totality of their lifetime within the time range 01 May 2010 through 31 December 2018; thus ARs which were already present on the Sun prior to 01 May 2010 or continued their presence on the Sun after 31 December 2018 are not included in this dataset.  NOAA ARs 11160, 11171, 12623, and 12705 never developed sunspots and thus contribute no images to this dataset.  Additionally, NOAA ARs 11190, 11493, 11494, 11496, 11501, 11503, 12472, 12473, and 12570 are not included in this dataset since they appeared during times when the SDO satellite was missing fine guidance~\cite{hmi_coverage} and thus the location of the ARs could not be accurately tracked.  (More specifically, a reference time (\url{http://jsoc.stanford.edu/doxygen_html/im__patch_8c-source.html}) is specified for the AR corresponding to the time that AR will be at disk center (\url{http://jsoc.stanford.edu/doxygen_html/libs_2astro_2heliographic__coords_8c-source.html}) and no data records are returned if there are no valid data within a four hour window of that reference time.)  The entire image set (i.e., the 1,372,004 \verb+.fits+ images) comprises 537 GB.  We also provide a pre-configured AR dataset of \verb+.fits+ images and corresponding flare labels, described below, which comprises 375 GB and a reduced size (spatially and bit-depth) dataset of \verb+.png+ images and corresponding flare labels, also described below, which comprises 15 GB.

\subsection{Parsing the Solar Region Summaries for Active Regions}
We used the NOAA SWPC SRS (\url{ftp://ftp.swpc.noaa.gov/pub/warehouse/}) to determine the dates a NOAA AR is visible on disk.  The SRS are downloaded as one \verb+.txt+ file per day.  We used \verb+Part I+ data in the SRS which detail those active regions with associated sunspot structures~\cite{SRS_readme}.  For each NOAA AR appearing in SRS \verb+Part I+, we store the NOAA AR number, the date the AR first appears in the SRS, and accumulate the total number of days the same AR appears in the SRS.  We store these data in a comma separated text file \verb+ARList.txt+ where each line is of the format \verb+NNNN,YYYYMMDD,X+, where \verb+NNNN+ is the four digit NOAA AR number, \verb+YYYYMMDD+ is the initial date of appearance, and \verb+X+ is an integer number of days.  The \verb+AR_List.txt+ file used to download the image set described here is provided as part of the GitHub repository at~\cite{ar_flares_github}.

\subsection{Downloading the Magnetograms for Active Regions}
The text file \verb+AR_List.txt+ as described above is used to specify an appropriate date range to download the HMI magnetograms centered on a given AR.  We request HMI magnetograms beginning at time 00:00:00 on the first day the AR appeared through 00:00:00 on the first day the AR disappeared.  While there are modules to access SDO data for python (e.g., sunpy~\cite{sunpy}) and IDL (e.g., SolarSoft~\cite{solarsoft}) without navigating the JSOC webpage, the ability to extract and track a cutout around a NOAA AR does not appear to be accessible through any means other than the website.  In order to automate this process to download the 1,655 ARs, we wrote a python script to interact with the webpage using the selenium package~\cite{selenium} and geckodriver~\cite{geckodriver} for Mozilla's firefox web browser.  

We provide this code as part of the GitHub repository~\cite{ar_flares_github}, but note that the code will break if any of the underlying html code on the JSOC website changes.  Since the JSOC driver code is fragile, we describe in detail the process of interacting with the JSOC Data Export webpage to download a single AR of data here.  Readers who are interested in using the curated datasets~\cite{preconfigured_dataset,reduced_dataset,extra_images_dataset} described in this paper can skip to the next subsection.  Readers who are interested in downloading a custom dataset from the JSCO Data Export webpage may be interested in the process described here.  This process assumes that the SWPC SRS have been parsed as in the previous section to determine the beginning date and number of days the AR is on disk.  
\begin{itemize}
    \item Navigate to the JSOC Data Export tool: \url{http://jsoc.stanford.edu/ajax/exportdata.html} 
    \item In the \verb+RecordSet+ field enter the data locator in the form
    \begin{center}\vspace{-5pt}
      \verb+hmi.M_720s[date1_time1_TAI-date2_time2_TAI][?quality>=0?]+ 
    \end{center}\vspace{-5pt}
    where dates and times are in the format \verb+YYYY.MM.DD_HH:MM:SS+, \verb+TAI+ is the designation for international atomic time used by SDO, and the \verb+quality+ keyword specifies a search only for observables that were created.  Press enter and the \verb+Record Count+ field will change to the total number of images spanned by the requested time period.  There should be approximately 120 images per day requested.  
    \item Using the \verb+Method+ dropdown menu, select \verb+url-tar+.
    \item Check the \verb+Enable Processing+ checkbox which will result in the appearance of several additional check boxes. 
    \item Check the \verb+im_patch+ checkbox which will result in the appearance of an \verb+Image Patch Extract+ box.
    \item In the \verb+Image Patch Extract+ box:
	    \begin{itemize}
	        \item Ensure \verb+Tracking+ is checked in the \verb+options+ row.
            \item Specify the \verb+NOAA AR number+ in the \verb+options+ row as a four or five digit number.  Press enter and the \verb+T_REF+, \verb+X+, and \verb+Y+ fields will populate with reference time and location information for the AR.    
            If the four digit truncated NOAA AR number 
            is entered, the field automatically changes to the corresponding five digit number. 
            \item Verify \verb+T_START+ and \verb+T_STOP+ match the dates given in the \verb+RecordSet+ field.
            \item Verify \verb+Cadence+ matches the cadence set in the \verb+RecordSet+ field.
            \item Verify \verb+BoxUnits+ is set to \verb+pixels+. 
            \item Set \verb+Width+ and \verb+Height+ to \verb+600+ each.
            \item Click the \verb+Check Params+ button which will change the adjacent text field from \verb+Not Ready+ to \verb+OK to submit+
	    \end{itemize}
	\item Verify \verb+Protocol+ is set to \verb+FITS+.
    \item Enter the user's email (to which notification will be sent when the data is ready to be downloaded) in \verb+Notify+ field and user's name in \verb+Requestor+ field.  The user's email must match a registered user (see also next bullet).
    \item Click \verb+Check params for export+ and the \verb+Not Ready To Submit+ button will change to a \verb+Submit Export+ \verb+Request+ button. If the email entered in the \verb+Notify+ field is not registered, a message will appear specifying that the user should respond to an email from JSOC within 15 minutes to register their email.  An email will be sent from jsoc@sun.Stanford.EDU with subject ``CONFIRM EXPORT ADDRESS'' with further instructions.  In short, a simple response to that email will register the user after which the user should receive a second email with subject ``EXPORT ADDRESS REGISTERED.''  After this initial registration process, the user will need to click on the \verb+Check params for export+ button again.  This registration process will need to be completed only once per user.
    \item Click \verb+Submit Export Request+ at which point the \verb+RequestID+ field will be populated with a string used to identify the data request %(\verb+JSOC_20200804+ \verb+_1553+ for this example).  
    There may be few second delay before the \verb+RequestID+ field will populate.
    \item At the bottom of the page in the \verb+JSOC Data Export Status and Re-+ \verb+trieval+ section, verify \verb+RequestID+ matches the above given \verb+RequestID+.
	\item Periodically click \verb+Submit Status Request+ until the \verb+Status+ field becomes \verb+Data Ready+.  The \verb+Status+ may say \verb+Bad Request Status+ for the first few clicks of \verb+Submit Status Request+; continue to click the same button until a request time is displayed in the \verb+Status+ field.
	\item When the \verb+Status+ field becomes \verb+Data Ready+, click on the link provided in the \verb+TarFile Location+ field to download the requested data.
\end{itemize}

\subsection{Parsing the Event Reports for Active Regions}
Using the SWPC Event Reports (ER)~\cite{Events_readme} we parsed the text data for \verb+XRA+ events in the \verb+Type+ column (corresponding to x-ray events detected by the Geospatial Operational Environmental Satellite (GOES) spacecraft) with an associated number in the \verb+REG#+ column (corresponding to a NOAA AR number).  For those x-ray events associated with a NOAA AR, we additionally parsed the ER for the peak flare time (\verb+Max+ column), and flare size (\verb+Particulars+ column). We store these data in a comma separated text file \verb+Event_List.txt+ where each line is of the format \verb+YYYY MM DD,HHMM,NNNN,KX.X+ where \verb+YYYY MM DD+ is the date, \verb+HHMM+ is the time, \verb+NNNN+ is the four-digit NOAA AR number, and \verb+KX.X+ is the GOES flare class (e.g., \verb+C1.0+ or \verb+X10.1+)~\cite{flare_class}.  The \verb+Event_List.txt+ file for this dataset is provided as part of the image set at~\cite{extra_images_dataset}.

\subsection{Customizing the Dataset}
In this section we provide details on the postprocessing of the dataset according to AR location and flaring behavior.  We provide a preconfigured dataset consisting of AR magnetograms within $\pm60^\circ$ latitude and longitude, containing zero NaN pixels, and labeled according to flaring behavior within 24 hours and at a flare size greater than C1.0.  As described above, we download magnetogram images for NOAA ARs for the duration of their appearance on the solar disk; hereafter, we refer to this as the ``image set'' to distinguish it from the ``AR dataset.''  The preconfigured AR dataset (described below) is available at~\cite{preconfigured_dataset} and a reduced resolution preconfigured AR dataset (described below) is available at~\cite{reduced_dataset}.  The image set can be acquired by combining the preconfigured AR dataset~\cite{preconfigured_dataset} and the extra images dataset~\cite{extra_images_dataset} which contains those images removed in the preconfiguration process.

\subsubsection{Filtering Data by Latitude, Longitude and Not-a-Number (NaN) Pixels)}
\begin{figure}
    \centering
    \includegraphics[width=0.7\textwidth]{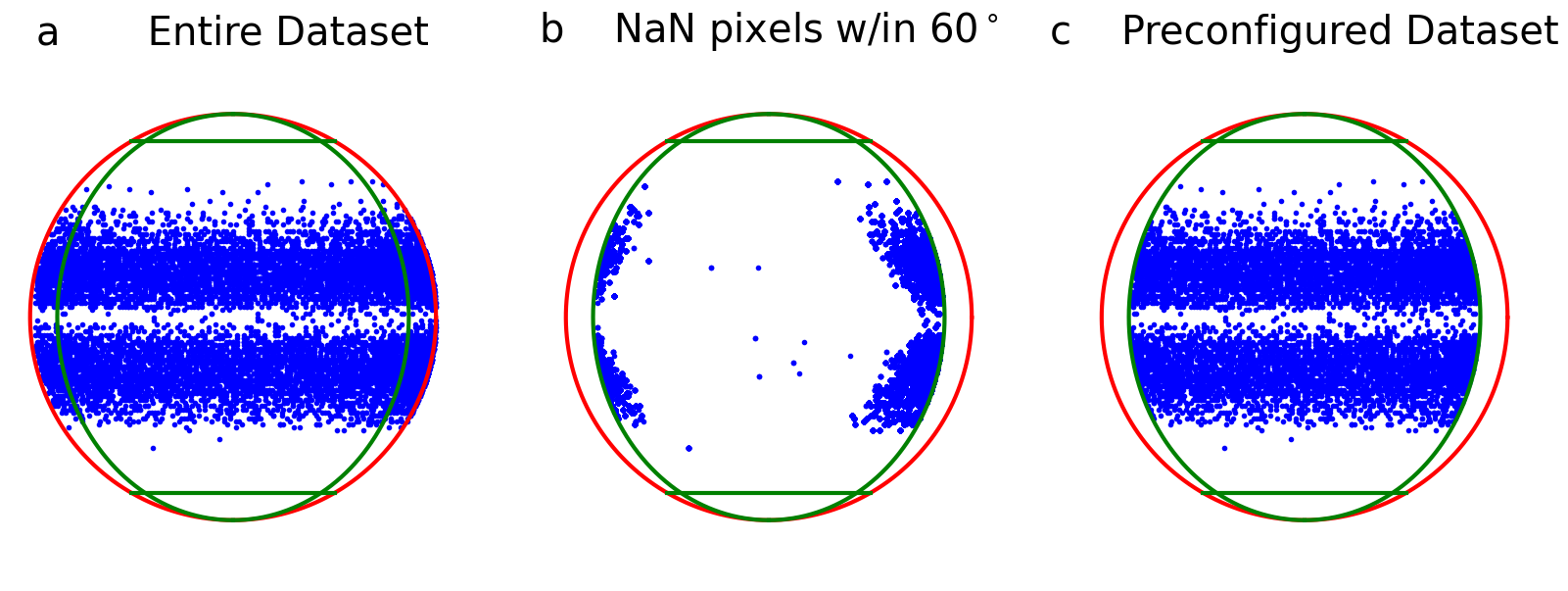}
    \caption{Latitude and longitude of AR images.  The red circle denotes the solar radius and the green lines denote $\pm60^\circ$ latitude and longitude.  The blue dots denote the centroids of the ARs included in the respective datasets. a: Latitude and longitude of files for entire dataset (image set).  b: Latitude and longitude of files within $\pm60^\circ$ and $\ge1$ NaN pixels.  c: Latitude and longitude of files for preconfigured AR dataset.}
    \label{fig:latlon}
\end{figure}
\begin{figure}[t]
    \centering
    \includegraphics[width=0.7\textwidth]{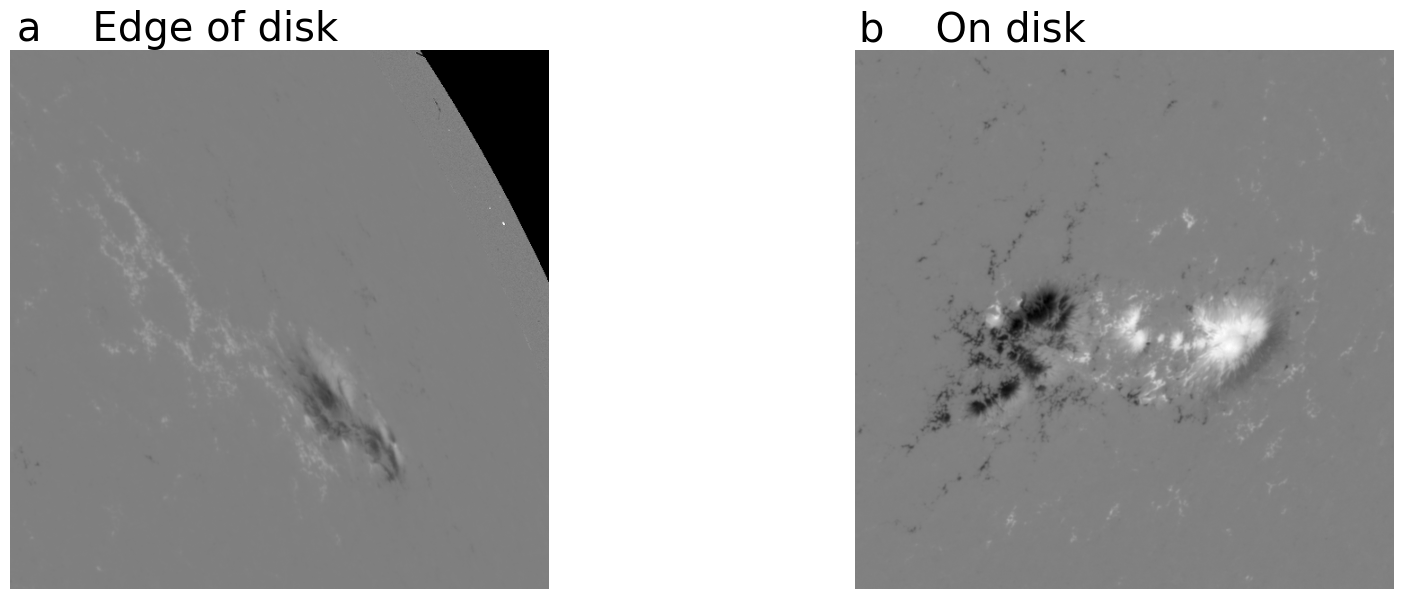}
    \caption{Examples of $600\times600$ pixel magnetogram images, including a disk-edge magnetogram and an on-disk magnetogram. a: Disk-edge magnetogram.  NOAA AR 1169, 2011 March 15, 12:00:00. b: On-disk magnetogram.  NOAA AR 2396, 2015 August 11, 00:00:00.}
    \label{fig:ex_mags}
\end{figure}

  Figure~\ref{fig:latlon}a shows a scatter plot of the latitude and longitude of the AR centroids for the image set.  Some of these images, however, are near the edge of the solar disk and parts of the image capture data from off the solar disk (see Figure~\ref{fig:ex_mags}a). These disk-edge images may contain nonsensical magnetic measurements or NaN values.  Furthermore, since the HMI magnetograms are line-of-sight (LOS), edge-of-disk images are affected by larger projection effects.  These projection effects depend not only on the viewing angle but also on the specific geometry of the magnetic field, with deviations from radial in regions of stronger magnetic field introducing larger projection errors~\cite{leka2017}.  In this dataset, we do not implement any correction for projection effects, e.g., those in~\cite{leka2017}, but do provide a means for the user to configure a dataset by restricting the resulting images to reside within latitude and longitude bounds to limit the errors introduced by projection effects.  We further note that the user could apply additional preprocessing methods to any of the image set images. 

We use the SRS to determine the latitude and longitude for an AR on a given date, noting that the latitude and longitude are provided in the SRS at a daily cadence.  Thus, we may exclude some images near the east limb that are just outside of the longitude threshold and rotate into a valid range throughout the day.  Similarly, we may include some images near the west limb that are just inside the longitude threshold and rotate out of the valid range throughout the day.  Using the daily latitude and longitude provided in the SRS files, we include in the preconfigured AR dataset all images with an AR centroid within $\pm60^\circ$ latitude and longitude (similar to those data in~\cite{algraibah2015,boucheron2015}).  A total of 313,601 files, comprising 22.9\% of the entire dataset, are excluded from the preconfigured AR dataset based on latitude and longitude; a total of 85 ARs are excluded entirely based on these criteria.

Due to the constant $600\times600$ pixel window of the images, ARs further from the equator may still contain off-disk data and we additionally exclude any image containing any NaN values, an additional 108,356 files and 7.9\% of the entire dataset.  The majority of these images with NaN values contain a small portion of the disk edge, but there are some images with spurious NaN values from various latitudes and longitudes.  Figure~\ref{fig:latlon}b shows a scatter plot of those ARs within $\pm60^\circ$ latitude and longitude which contained at least one NaN pixel.  We note that the majority of these images are near the disk edge, with a higher number of these images clustered near the west limb as compared to the east limb.  This is consistent with the expectation that active regions on the west limb will be rotating closer to the disk edge throughout the day.  

In total, between the latitude/longitude filtering and the NaN filtering, we exclude 421,957 images, comprising 30.8\% of the entire dataset, from the preconfigured dataset.  This results in a preconfigured dataset consisting of 950,047 on-disk HMI images (see Figure~\ref{fig:ex_mags}b) within a range of latitudes and longitudes (see Figure~\ref{fig:latlon}c).  We provide the 950,047 images as part of the preconfigured AR dataset~\cite{preconfigured_dataset} and the reduced resolution dataset~\cite{reduced_dataset}.

\subsubsection{Assigning Flare Labels to Images}
In order to use the dataset for supervised classification or regression, each image in the AR dataset needs a corresponding label specifying whether that image is associated with a flare.  We provide a label indicating the flare size (as a string of GOES class, e.g., \verb+'C1.0'+) for images associated with flares or \verb+'0'+ for images associated with non-flaring behavior.  The user can configure the minimum flare size as well as the temporal flare prediction window; any images within the prediction window leading up to a flare are associated with that flare.  For those ARs that flare multiple times within the flare prediction window, images are assigned a class associated with the largest size flare, consistent with~\cite{boucheron2015}. 

\begin{figure}
    \centering
    \includegraphics[width=0.9\textwidth]{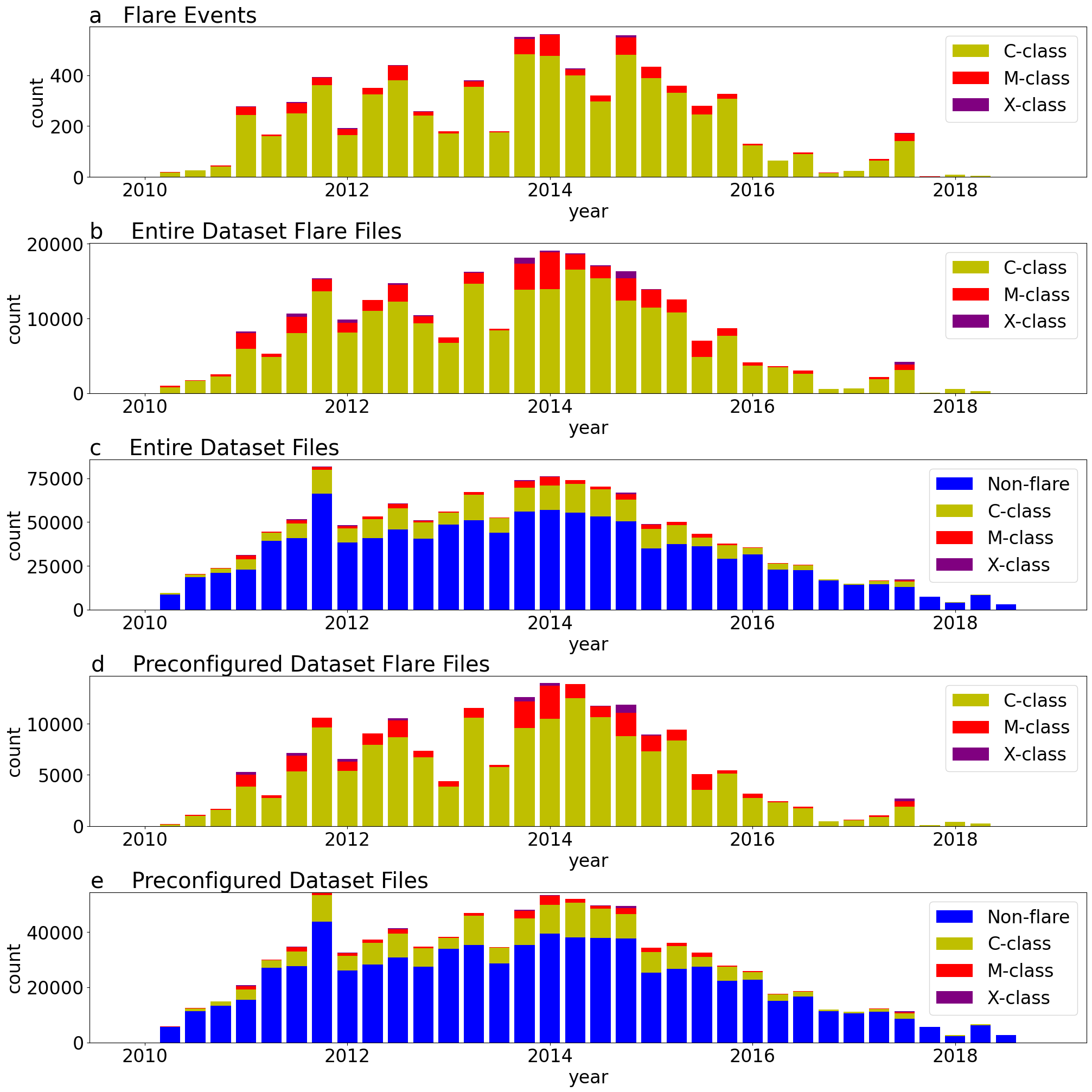}
    \caption{Count of events or files for different flaring behavior versus quarter. a: Count of flare events in the entire dataset. b: Flare file count for the entire dataset.  c: Flare and non-flare file count for the entire dataset.  d: Flare file count for the preconfigured dataset.  e: Flare and non-flare file count for the preconfigured dataset.}
    \label{fig:flarecounts}
\end{figure}

Figure~\ref{fig:flarecounts}a shows a plot of the number of C-, M-, and X-class flares during the timespan of this dataset, while Figures~\ref{fig:flarecounts}b and~\ref{fig:flarecounts}c show counts of images associated with flaring behavior for a 24 hour flare prediction window for the entire dataset.  We notice very similar trends in the count of flare events (Figure~\ref{fig:flarecounts}a) and the count of files associated with a flare (Figure~\ref{fig:flarecounts}b).  This indicates that the entire dataset has well-sampled the flaring behavior of the Sun over this time period.

In order to assign labels to the AR dataset images, we loop over each event in \verb+Event_List.txt+ and assign a label of the GOES flare size for all images of the AR within 24 hours of the peak flare time for any flare size that satisifes the user-specified minimum flare size.  After assigning flaring images for all events in \verb+Event_List.txt+, all remaining images are labeled \verb+'0'+ to denote non-flaring images.  The flare labels are stored in a file \verb+KX.X_Hhr_Labels.txt+ file where \verb+KX.X+ is the user-specified minimum flare size, e.g., \verb+C1.0+, and \verb+H+ is the user-specified prediction window in hours, e.g., \verb+24+.  Each line in the flare labels file is of the form \verb+filename,label+ where \verb+filename+ is the base filename and \verb+label+ is the label (flare size for flaring and \verb+'0'+ for non-flaring).

For the preconfigured AR dataset, we specify a 24 hour prediction window and a minimum flare size of C1.0.  We provide the \verb+C1.0_24hr_Labels.txt+ file as part of the preconfigured AR dataset~\cite{preconfigured_dataset} and the \verb+C1.0_24hr_png_Labels.txt+ file as part of the preconfigured reduced resolution AR datset~\cite{reduced_dataset}, both of which contain 190,582 flaring images and 759,465 non-flaring images.  Figures~\ref{fig:flarecounts}d and~\ref{fig:flarecounts}e show plots of images associated with flaring behavior for the preconfigured AR dataset.  We notice very similar trends in the count of flare events for the entire dataset (Figures~\ref{fig:flarecounts}b and~\ref{fig:flarecounts}c) and in the preconfigured AR dataset (Figures~\ref{fig:flarecounts}d and~\ref{fig:flarecounts}e).  This indicates that the configuration of the preconfigured AR dataset based on latitude, longitude, and presence of NaNs in the images has not significantly altered the distribution of flare classes.  

\subsection{Dataset Partitions}
To facilitate comparison between flare prediction methods, we have partitioned the dataset into training, validation, and testing sets.  To this end, we randomly selected 10\% of the ARs to set aside for validation purposes (e.g., tuning of algorithm parameters), an additional 10\% of the ARs for testing purposes, and the remaining 80\% for training purposes.  We note that the initial random assignment of ARs resulted in a validation set with different classification performance, specifically a higher true positive rate (TPR), on several classification tasks.  Further investigation found that the validation set had a higher proportion of ARs with very high TPR.  Randomly re-assigning seven ARs with TPR$>$0.90 from validation to test and a random seven ARs with TPR$<$0.90 from test to validation resulted in more similar performance between test and validation.  There are 157 ARs and 94,757 images in the test data, 157 ARs and 95,933 images in the validation data, and 1,256 ARs and 759,357 images in the training data.  Lists of the ARs included in each of the three sets are provided in files \verb+List_of_AR_in_Train_Data_by_AR.csv+, \verb+List_of_AR_in_Validation_Data_by_ARcsv+, and \verb+List_of_AR_in_Test_Data_by_AR.csv+ as part of the dataset repositories~\cite{preconfigured_dataset,reduced_dataset}.

\subsection{Reduced Resolution Dataset}
We have created a reduced resolution dataset at a spatial resolution of $224\times224$ pixels and bit depth of 8 bits.  This reduced resolution dataset contains images in \verb+.png+ format which are more readily ingested by standard image processing libraries and at bit-depth and spatial resolution compatible with typical convolutional neural network (CNN) architectures.  The spatial resolution is reduced using the \verb+transform.resize+ command in scikit-image with parameters \verb+order=1+, \verb+mode='reflect'+, \verb+clip=True+, \verb+preserve_range=True+, \verb+anti_aliasing=True+.  The bit depth of the resized images are reduced by 1) offsetting the intensities by 2550, 2) clipping to the range $[0,5100]$, 3) scaling to the range $[0,255]$, and 4) rounding to the nearest integer:
\begin{equation}
    I_{8} = \left[\text{Min}\text{Max}(0,5100,I+2550)\frac{255}{5100}\right],
\end{equation}
where $I_8$ is the image in unsigned 8-bit integer bit-depth resolution, $I$ is the input image, $\text{Min}\text{Max}(mn,mx,x)$ denotes a clipping of $x$ to the range $[mn,mx]$, and $[\cdot]$ denotes a round operation.  It should be noted that this reduction in bit depth results in an error due to both the clipping operation and the scaling operation. The clipping operation affects only 2e-4\% of pixels in the entire dataset which originally corresponded to the largest flux values (positive and negative).  The scaling operation will result in a range of 20 G being mapped to the same intensity level with an error in the range $[-10,10]$ G which is on the order of the noise level of the HMI instrument~\cite{hmi_noise}.

\section{Data Records}
The preconfigured dataset~\cite{preconfigured_dataset}, reduced resolution dataset~\cite{reduced_dataset}, and extra images dataset~\cite{extra_images_dataset} each contain a directory structure \verb+Lat60_Lon60_Nans0+, \verb+Lat60_Lon60_Nans0_png_224+, and \verb+active_regions_extra+, respectively.  This directory structure contains the ARs in four digit directory names, e.g., \verb+1325+.  Each directory contains multiple magnetogram images in \verb+.fits+ format~\cite{preconfigured_dataset,extra_images_dataset} or \verb+.png+ format~\cite{reduced_dataset}.  The base filenames are defined with the format\\
\verb+hmi.M_720s.YYYYMMDD_HHMMSS_TAI.1.magnetogram+ as downloaded from JSOC.

The preconfigured dataset~\cite{preconfigured_dataset} and the reduced resolution dataset~\cite{reduced_dataset} additionally contain the following directory structures and files of use for classification and regression tasks.  In the following, the first filename corresponds to the preconfigured dataset~\cite{preconfigured_dataset} and the second filename corresponds to the reduced resolution dataset~\cite{reduced_dataset}; if only one filename is given, the filenames (and files) are identical between the two datasets.
\begin{itemize}
    \item  \verb+C1.0_24hr_Labels.txt+, \verb+C1.0_24hr_224_png_Labels.txt+: a file containing the labels for each of the images in the dataset.  The labels are formatted to provide both the regression and classification labels in a form that can be parsed for other applications.  Each line in the file is of the form \verb+filename,label+ where \verb+filename+ is the base filename in the image set and \verb+label+ is the label.  The label is formatted as a string \verb+KX.X+ for flaring regions, where \verb+K+ is the GOES flare class (\verb+C+, \verb+M+, or \verb+X+) and \verb+X.X+ is the size, e.g., \verb+4.7+.  Non-flaring regions are assigned a label of  \verb+'0'+.
    \item \verb+List_of_AR_in_Train_Data_by_AR.csv+, \verb+List_of_AR_in_Validation_Data_by_ARcsv+, and\\ \verb+List_of_AR_in_Test_Data_by_AR.csv+: files containing lists of NOAA ARs assigned to the training, validation, and test sets, respectively.  Each line in the files is of the format \verb+NNNN+, the four digit NOAA AR number.  Note--these lists are identical between the reduced resolution dataset and the full resolution dataset.
    \item \verb+Lat60_Lon60_Nans0_C1.0_24hr_features.csv+,\\
    \verb+Lat60_Lon60_Nans0_C1.0_24hr_png_224_features.csv+: a file with the 29 magnetic complexity features extracted from each of the reduced resolution images in the preconfigured dataset.  Each line of the file contains 32 comma separated values.  The first 29 values are the 29 magnetic complexity features as described below.  The last three values are the classification label (\verb+1+ or \verb+0+), regression label (flare size as as string \verb+KX.X+ or \verb+0+), and the base filename.  The regression label is formatted as a string \verb+KX.X+ for flaring regions, where \verb+K+ is the GOES flare class (\verb+C+, \verb+M+, or \verb+X+) and \verb+X.X+ is the size, e.g., \verb+4.7+.
    \item  (\verb+Train_Data_by_AR.csv+, \verb+Train_Data_by_AR_png_224.csv+), (\verb+Validation_Data_by_AR.csv+,\\
    \verb+Validation_Data_by_AR_png_224.csv+), (\verb+Test_Data_by_AR.csv+, \verb+Test_Data_by_AR_png_224.csv+): files with labels for each of the images in the preconfigured dataset formatted to provide classification labels in the format expected by a dataframe loader in tensorflow for the training, validation, and test sets, respectively.  Each line is of the form \verb+NNNN/filename,label+ where \verb+NNNN+ is the AR directory, \verb+filename+ is the base filename, and \verb+label+ is the classification label (\verb+1+ for flaring and \verb+0+ for nonflaring).
\end{itemize}

The extra images dataset~\cite{extra_images_dataset} contains a file \verb+EventList.txt+ which contains the list of events (flares) occurring within the timespan of the dataset.  Each line is of the format \verb+YYYY MM DD,HHMM,NNNN,KX.X+ where \verb+YYYY MM DD+ is the date, \verb+HHMM+ is the time, \verb+NNNN+ is the four-digit NOAA AR number, and \verb+KX.X+ is the GOES class (e.g., \verb+C1.0+ or \verb+X10.1+).

\section{Technical Validation}

In this section we describe two experiments that demonstrate the utility of the preconfigured AR dataset.  In the first, we implement a flare prediction method using magnetic complexity features and a support vector machine (SVM) classifier.  In the second, we provide preliminary results of a transfer learning approach for use of convolutional neural networks (CNNs) for flare prediction.

\subsection{Magnetic Complexity Features for Machine Learning}
\begin{figure}
    \centering
    \includegraphics[width=0.9\textwidth]{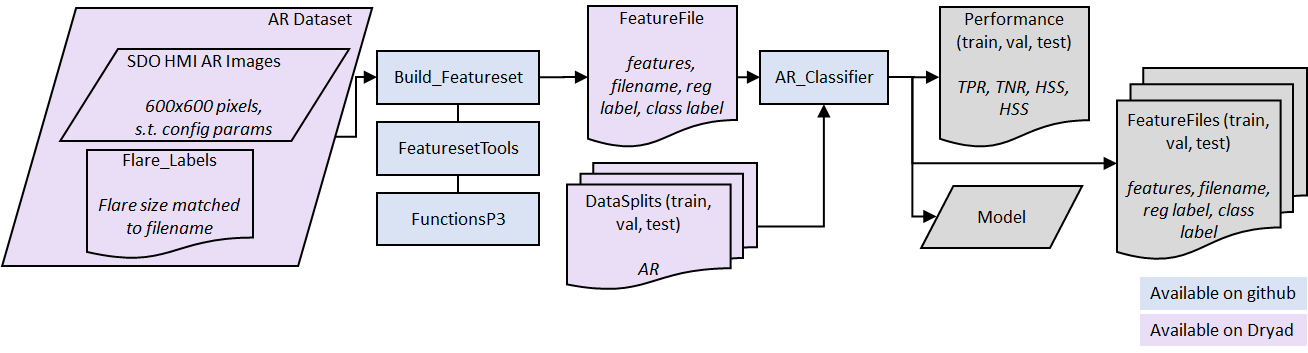}
    \caption{Flowchart of SVM classification of flare activity.  }
    \label{fig:SVM_flowchart}
\end{figure}

We extract 29 of the 38 magnetic complexity features of~\cite{algraibah2015} from each of the HMI magnetograms in the preconfigured AR dataset and use a support vector machine (SVM) to predict whether the AR will flare within the next 24 hours.  An overview of the SVM classification is shown in Figure~\ref{fig:SVM_flowchart}.  The methods presented in~\cite{algraibah2015} were applied to MDI magnetograms which have lower spatial resolution ($\sim2''\times2''$ pixels), and lower cadence (96 minutes) than the HMI dataset presented here ($\sim0.5''\times0.5''$ pixels and 12 minute cadence).  Due the lower cadence of the MDI magnetograms, the dataset was also much smaller, with approximately 260,000 total images.  The 9 flux evolution features from~\cite{algraibah2015} are omitted in this work: these features require a comparison between two images and therefore cannot be directly linked to a single image, the cadence of the HMI magnetograms is 12 minutes (as opposed to 96 minutes) leading to minimal evolution of an AR between images in this dataset, and the flux evolution features proved to be poor features for classifying ARs.

We provide the 29 magnetic features as part of the preconfigured AR dataset~\cite{preconfigured_dataset} and for the reduced resolution dataset~\cite{reduced_dataset} and the code to extract the magnetic features on GitHub at~\cite{ar_flares_github}.  These magnetic complexity features include 7 gradient features, 13 neutral line features, 5 wavelet features, and 4 flux features in the format of a \verb+.csv+ file. Each row in the \verb+.csv+ file represents an image in the dataset.  The first 29 columns are the 29 magnetic features.  The 30th column is the binary flare class (\verb+'1'+ or \verb+'0'+) and the 31st column is the flare size in terms of the GOES flare class (with a value of \verb+'0'+ representing no flare or a flare smaller than \verb+'C1.0'+).  The last column is the filename of the image corresponding to the magnetic features and flare class.

\begin{table}[t]
    \centering
    \begin{tabular}{l|l|l|l|l|l}
        \textbf{Method} & \textbf{Dataset} & \textbf{TPR} & \textbf{TNR} & \textbf{HSS} & \textbf{TSS}\\\hline
        \multirow{2}{*}{SVM} & Full Resolution & 0.7484 & 0.7791 & 0.4485 & 0.5275 \\
         & Reduced Resolution & 0.7884 & 0.7464 & 0.4350 & 0.5348%\\\hline
        % \multirow{2}{*}{VGG} & Full Resolution & 0.6293 & 0.8828 & 0.4872 & 0.5120 \\
        %  & Reduced Resolution & 0.7765 & 0.7569 & 0.2995 & 0.5334
    \end{tabular}
    \caption{SVM performance on the test dataset for the full resolution and reduced resolution datasets.}
    \label{tab:SVM_performance}
\end{table}

An SVM classifier is trained on the training set using the \verb+SVC+ function from scikit-learn; this code is also available on GitHub at~\cite{ar_flares_github}.  All parameters were left as the default (\verb+C=1.0+, \verb+shrinking=True+, \verb+probability=False+, \verb+tol=0.001+, \verb+decision_function_shape='ovr'+, \verb+break_ties=False+, \verb+random_state=None+) with the exception of the \verb+kernel+ parameter which was set to \verb+'linear'+ and the \verb+class_weight+ parameter which was set to \verb+'balanced'+ to account for the imbalanced nature of this dataset. This experiment is intended as a validation of the use of the datasets for classical machine learning methods.  As such, we have not optimized the kernel or parameters of the classifier. Performance metrics are evaluated on the test set and are summarized in Table~\ref{tab:SVM_performance}.  The performance metrics considered are the true positive rate (TPR), true negative rate (TNR), Heidke skill score (HSS), and true skill statistic (TSS) as defined in~\cite{algraibah2015}.  As a comparison the work in~\cite{algraibah2015} achieved a TPR of 0.81, TNR 0.70, HSS 0.39, and TSS 0.51.  Given that work was applied to a different dataset from a different instrument, we find the results here comparable to that work and a validation of the utility of this dataset for flare prediction.  We also note that the comparable performance between the full and reduced resolution data indicates that the reduced resolution dataset has retained the vast majority of the information needed for this classification problem.  We note, however, that other machine learning tasks may benefit from the increased spatial or bit depth resolution of the full resolution dataset.

\subsection{Deep Learning}
\begin{figure}
    \centering
    \includegraphics[width=0.6\textwidth]{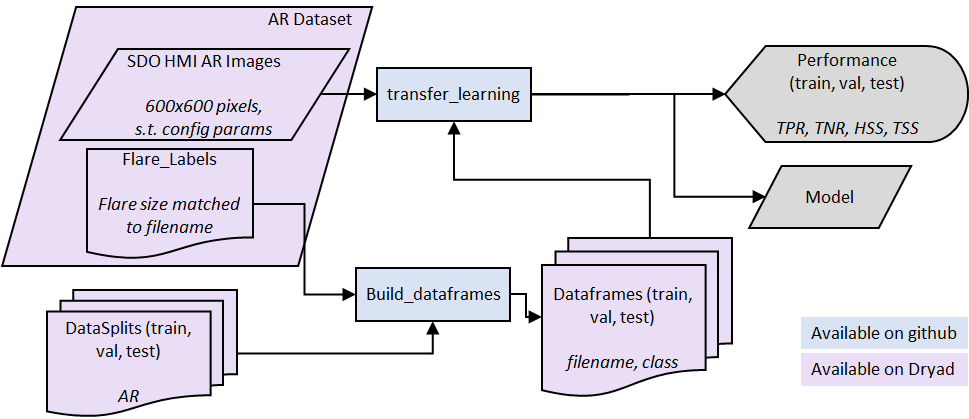}
    \caption{Flowchart of VGG classification of flare activity.  }
    \label{fig:VGG_flowchart}
\end{figure}

We perform transfer learning on the VGG16 CNN~\cite{vgg}, pretrained on ImageNet using the \verb+tensorflow.keras+~\cite{tensorflow} VGG model.  An overview of the SVM classification is shown in Figure~\ref{fig:VGG_flowchart} and code is available on GitHub at~\cite{ar_flares_github}.  We replace the final fully connected layer (originally $4096\times1000$) with a $4096\times2$ layer with softmax activation.  In training, we freeze all layers except that final fully connected layer.  For the full resolution data in \verb+.fits+ format, a custom data generator was written since the \verb+.fits+ format is not one that \verb+tensorflow+ can handle natively.  Within that data generator, the images are resized to the expected spatial dimensions ($224\times224$ pixels) using the \verb+skimage.transforms.resize+ command with options \verb+order=1+, \verb+mode='reflect'+, \verb+clip=True+, \verb+preserve_range=True+, and \verb+anti_aliasing=True+ and to the expected intensity range by linearly scaling the full range of the data $[-5978.7,5978.7]$ to $[0,255]$.  The images are then preprocessed with the built-in \verb+preprocess_input+ function as part of the \verb+tensorflow.keras+ VGG model.  For the reduced resolution dataset, the \verb+flow_from_dataframe+ method is used along with the VGG \verb+preprocess_input+ preprocessing.  Both data generators use a batch size of 64.  For training, we used the adam optimizer with options \verb+learning_rate=0.001+, \verb+beta_1=0.9+, \verb+beta_2=0.999+, \verb+epsilon=1e-07+, and \verb+amsgrad=False+ and the categorical cross-entropy loss.  The networks are trained for 5 epochs with the \verb+class_weight+ parameter set to 1 for the majority (non-flare) class and $N_n/N_f$ for the minority (flare) class, where $N_n$ is the number of nonflaring examples and $N_f$ is the number of flaring examples.  We wrote custom \verb+tensorflow.keras+ metrics to track the TPR, TNR, HSS, and TSS throughout the training process.  This experiment is intended as a validation of the use of the datasets for deep learning methods.  As such, we have not optimized the architecture, which layers are frozen, or optimizer parameters.  The best model was chosen as the epoch with the maximum validation TSS.  Performance on the test data is summarized in Table~\ref{tab:VGG_performance}.  We see scores commensurate with the SVM performance, indicating the validity of this dataset in deep learning methods.

\begin{table}[t]
    \centering
    \begin{tabular}{l|l|l|l|l|l}
        \textbf{Method} & \textbf{Dataset} & \textbf{TPR} & \textbf{TNR} & \textbf{HSS} & \textbf{TSS}\\\hline
        % \multirow{2}{*}{SVM} & Full Resolution & 0.7483 & 0.7791 & 0.4485 & 0.5275 \\
        %  & Reduced Resolution & 0.7884 & 0.7464 & 0.4350 & 0.5348\\\hline
        \multirow{2}{*}{VGG} & Full Resolution & 0.6952 & 0.8142 & 0.4567 & 0.5094 \\
         & Reduced Resolution & 0.7344 & 0.7980 & 0.4636 & 0.5325
    \end{tabular}
    \caption{VGG performance on the test dataset for the full resolution and reduced resolution datasets.}
    \label{tab:VGG_performance}
\end{table}

\section*{Usage Notes}
Further details on usage of the datasets can be found as part of the dataset repository documentation for the preconfigured dataset~\cite{preconfigured_dataset}, reduced resolution dataset~\cite{reduced_dataset} and extra images dataset~\cite{extra_images_dataset}.  Further details on usage of the code for configuration of the datasets and classification can be found as part of the GitHub repository at \url{https://github.com/DuckDuckPig/AR-flares.git}.

\section*{Code availability}
All code used to generate and manipulate the dataset, as well as code used in the Technical Validation is available at the GitHub repository \url{https://github.com/DuckDuckPig/AR-flares.git}.  Further details and documentation regarding code usage are included therein.

% \section*{Author contributions statement}
% T.V., L.E.B.~performed the data collection.  T.V., L.E.B.~curated the dataset.  E.W., J.A.G., L.E.B.~performed the SVM prediction simulations. T.V., L.E.B.~performed the deep learning prediction simulations.   All authors wrote and revised the paper.

% \section*{Competing interests} 
% The authors declare no competing interests.

\bibliographystyle{unsrt}
\bibliography{main}

\begin{thebibliography}{10}

\bibitem{pesnell2012}
W~Dean Pesnell, B~J Thompson, and PC~Chamberlin.
\newblock The solar dynamics observatory ({SDO}).
\newblock {\em Solar Physics}, 275(1-2):3--15, 2012.

\bibitem{lemen2012}
James~R Lemen, David~J Akin, Paul~F Boerner, Catherine Chou, Jerry~F Drake,
  Dexter~W Duncan, Christopher~G Edwards, Frank~M Friedlaender, Gary~F Heyman,
  Neal~E Hurlburt, et~al.
\newblock The atmospheric imaging assembly ({AIA}) on the solar dynamics
  observatory ({SDO}).
\newblock {\em Solar Physics}, 275(1-2):17--40, 2012.

\bibitem{woods2012}
TN~Woods, FG~Eparvier, R~Hock, AR~Jones, D~Woodraska, D~Judge, L~Didkovsky,
  J~Lean, J~Mariska, H~Warren, et~al.
\newblock Extreme ultraviolet variability experiment ({EVE}) on the solar
  dynamics observatory ({SDO}): Overview of science objectives, instrument
  design, data products, and model developments.
\newblock {\em Solar Physics}, 275(1-2):115--143, 2012.

\bibitem{scherrer2012}
Philip~Hanby Scherrer, Jothiram Schou, RI~Bush, AG~Kosovichev, RS~Bogart,
  JT~Hoeksema, Y~Liu, TL~Duvall, J~Zhao, CJ~Schrijver, et~al.
\newblock The helioseismic and magnetic imager ({HMI}) investigation for the
  solar dynamics observatory ({SDO}).
\newblock {\em Solar Physics}, 275(1-2):207--227, 2012.

\bibitem{algraibah2015}
Amani Al-Ghraibah, LE~Boucheron, and R.~T.~J. McAteer.
\newblock An automated classification approach to ranking photospheric proxies
  of magnetic energy build-up.
\newblock {\em Astronomy \& Astrophysics}, 579:A64, 2015.

\bibitem{boucheron2015}
Laura~E Boucheron, Amani Al-Ghraibah, and R.~T.~James McAteer.
\newblock Prediction of solar flare size and time-to-flare using support vector
  machine regression.
\newblock {\em The Astrophysical Journal}, 812(1):51, 2015.

\bibitem{canfield2000}
Richard~C Canfield and P~Murdin.
\newblock Solar active regions.
\newblock {\em Encyclopedia of Astronomy and Astrophysics}, 3:2457--2462, 2000.

\bibitem{hmi_coverage}
{HMI} observables data coverage plots.
\newblock \url{http://jsoc.stanford.edu/data/cov.html}.

\bibitem{SRS_readme}
{Space Weather Prediction Center SOLAR REGION SUMMARY (SRS)}.
\newblock \url{ftp://ftp.swpc.noaa.gov/pub/forecasts/SRS/README}.

\bibitem{ar_flares_github}
{AR-flares}.
\newblock \emph{GitHub} \url{https://github.com/DuckDuckPig/AR-flares}.

\bibitem{sunpy}
{The SunPy Community}, Will~T. Barnes, Monica~G. Bobra, Steven~D. Christe,
  Nabil Freij, Laura~A. Hayes, Jack Ireland, Stuart Mumford, David
  Perez-Suarez, Daniel~F. Ryan, Albert~Y. Shih, Prateek Chanda, Kolja
  Glogowski, Russell Hewett, V.~Keith Hughitt, Andrew Hill, Kaustubh Hiware,
  Andrew Inglis, Michael S.~F. Kirk, Sudarshan Konge, James~Paul Mason,
  Shane~Anthony Maloney, Sophie~A. Murray, Asish Panda, Jongyeob Park, Tiago
  M.~D. Pereira, Kevin Reardon, Sabrina Savage, Brigitta~M. Sipőcz, David
  Stansby, Yash Jain, Garrison Taylor, Tannmay Yadav, Rajul, and Trung~Kien
  Dang.
\newblock The sunpy project: Open source development and status of the version
  1.0 core package.
\newblock {\em The Astrophysical Journal}, 890:68--, 2020.

\bibitem{solarsoft}
{SolarSoft}.
\newblock \url{http://www.lmsal.com/solarsoft/sswdoc/index_menu.html}.

\bibitem{selenium}
selenium 3.141.0.
\newblock \url{https://pypi.org/project/selenium/}.

\bibitem{geckodriver}
mozilla/geckodriver.
\newblock \url{https://github.com/mozilla/geckodriver}.

\bibitem{preconfigured_dataset}
L~E Boucheron, T~Vincent, J~A Grajeda, and Wuest Ellery.
\newblock Active region magnetograms for solar flare prediction: Full
  resolution dataset.
\newblock \emph{Dryad} \url{https://doi.org/10.5061/dryad.dv41ns23n}, 2023.

\bibitem{reduced_dataset}
L~E Boucheron, T~Vincent, J~A Grajeda, and Wuest Ellery.
\newblock Active region magnetograms for solar flare prediction: Reduced
  resolution dataset.
\newblock \emph{Dryad} \url{https://doi.org/10.5061/dryad.jq2bvq898}, 2023.

\bibitem{extra_images_dataset}
L~E Boucheron, T~Vincent, J~A Grajeda, and Wuest Ellery.
\newblock Active region magnetograms for solar flare prediction: Extra images
  dataset.
\newblock \emph{Dryad} \url{https://doi.org/10.5061/dryad.qjq2bvqmj}, 2023.

\bibitem{Events_readme}
{EDITED SOLAR EVENTS LISTS}.
\newblock \url{ftp://ftp.swpc.noaa.gov/pub/indices/events/README}.

\bibitem{flare_class}
Karen~C. Fox.
\newblock Classifying solar eruptions.
\newblock
  \url{https://www.nasa.gov/mission_pages/sunearth/news/classify-flares.html},
  2012.

\bibitem{leka2017}
KD~Leka, G~Barnes, and EL~Wagner.
\newblock Evaluating (and improving) estimates of the solar radial magnetic
  field component from line-of-sight magnetograms.
\newblock {\em Solar Physics}, 292(2):36, 2017.

\bibitem{hmi_noise}
Sebastien Couvidat, Jesper Schou, J~Todd Hoeksema, Rick~S Bogart, Rock~I Bush,
  Tom~L Duvall, Yang Liu, Aimee~A Norton, and Philip~H Scherrer.
\newblock Observables processing for the {H}elioseismic and {M}agnetic {I}mager
  instrument on the {S}olar {D}ynamics {O}bservatory.
\newblock {\em Solar Physics}, 291:1887--1938, 2016.

\bibitem{vgg}
Karen Simonyan and Andrew Zisserman.
\newblock Very deep convolutional networks for large-scale image recognition.
\newblock Preprint at \url{https://arxiv.org/abs/1409.1556}, 2014.

\bibitem{tensorflow}
Mart\'{i}n Abadi, Ashish Agarwal, Paul Barham, Eugene Brevdo, Zhifeng Chen,
  Craig Citro, Greg~S. Corrado, Andy Davis, Jeffrey Dean, Matthieu Devin,
  Sanjay Ghemawat, Ian Goodfellow, Andrew Harp, Geoffrey Irving, Michael Isard,
  Yangqing Jia, Rafal Jozefowicz, Lukasz Kaiser, Manjunath Kudlur, Josh
  Levenberg, Dandelion Man\'{e}, Rajat Monga, Sherry Moore, Derek Murray, Chris
  Olah, Mike Schuster, Jonathon Shlens, Benoit Steiner, Ilya Sutskever, Kunal
  Talwar, Paul Tucker, Vincent Vanhoucke, Vijay Vasudevan, Fernanda Vi\'{e}gas,
  Oriol Vinyals, Pete Warden, Martin Wattenberg, Martin Wicke, Yuan Yu, and
  Xiaoqiang Zheng.
\newblock {TensorFlow}: Large-scale machine learning on heterogeneous systems,
  2015.
\newblock Software available from \url{https://www.tensorflow.org}.

\end{thebibliography}

\end{document}